\documentclass[final,5p,times,twocolumn]{elsarticle}

\usepackage{amssymb}
\usepackage{amsmath}
\usepackage{lipsum}
\usepackage{braket}
\usepackage{hyperref}
\usepackage[normalem]{ulem}
\usepackage[x11names]{xcolor}
\definecolor{vdrgreen}{rgb}{0.0, 0.6, 0.0}

\definecolor{hanblue}{rgb}{0.27, 0.42, 0.81}
\definecolor{byzantium}{rgb}{0.44, 0.16, 0.39}

\def\cevns{CE$\nu$NS }

\journal{}

\begin{document}

\begin{frontmatter}

\title{Inelastic neutrino-nucleus scattering off $^{203/205}$Tl in terms of the nuclear recoil energy using a hybrid nuclear model}

\author[JYU]{M. Hellgren\corref{cor1}}
\ead{majokahe@jyu.fi}
\author[IFC]{D. K. Papoulias}
\author[JYU,CIFRA]{J. Suhonen}
\affiliation[JYU]{organization={University of Jyväskylä},
            addressline={P.O. Box 35}, 
            city={Jyväskylä},
            postcode={FI-40014},
            country={Finland}}
\affiliation[IFC]{organization={Instituto de Física Corpuscular},
            addressline={Parc Científic UV C/ Catedrático José Beltrán, 2}, 
            city={Paterna (Valencia)},
            postcode={E-46980}, 
            country={Spain}}
\affiliation[CIFRA]{organization={International Centre for Advanced Training and Research in Physics (CIFRA)},
            addressline={P.O. Box MG12}, 
            city={Bucharest-Magurele},
            postcode={077125},
            country={Romania}}
\cortext[cor1]{Corresponding author.}

\begin{abstract}
Nuclear structure calculations in the context of a novel hybrid nuclear model, combining the nuclear shell model and the microscopic quasiparticle-phonon model are presented. The predictivity of the hybrid model is tested by computing inelastic neutral-current neutrino-nucleus scattering cross sections off the stable thallium isotopes. The cross sections are presented in terms of the incoming neutrino energy, taking also into account the effect of nuclear recoil energy.  Also reported are the expected event rates assuming neutrinos emerging from pion-decay at rest and the diffuse supernova neutrino background. Regarding solar neutrino rates, new results are presented in the context of the hybrid model and compared with previously reported results based solely on nuclear shell model calculations, demonstrating the improved accuracy of the adopted hybrid model at higher neutrino energies.
\end{abstract}

\begin{keyword}
Neutrino-nucleus scattering \sep Hybrid model \sep Scattering cross sections \sep Stopped-pion neutrinos \sep CE$\nu$NS
\end{keyword}

\end{frontmatter}

\section{Introduction}
Neutrinos and their properties are a subject of high interest and constitute one of the most active fields of research in contemporary physics. Within the Standard Model (SM) they experience only weak and gravitational interactions, of which only the former is relevant in terrestrial experiments aiming to detect them through their scattering off ordinary matter, such as atomic nuclei. These types of experiments are thus notably challenging due to the extremely small cross sections involved, with the research of neutrino-nucleus scattering being of central importance. There are a large number of past and actively running experiments, and the demand for theoretical neutrino-nucleus scattering studies to complement these is strong. In addition, relatively recent experimental results on the detection of the coherent elastic neutrino-nucleus scattering (CE$\nu$NS) process using stopped-pion (COHERENT~\cite{COHERENT:2017ipa,COHERENT:2020iec,COHERENT:2021xmm,COHERENT:2024axu}), $^{8}$B solar neutrinos (XENONnT~\cite{XENON:2024ijk}, PandaX-4T~\cite{PandaX:2024muv}) and rector antineutrino (Dresden-II~\cite{Colaresi:2022obx}, CONUS+~\cite{Ackermann:2025obx}) sources represent important milestones in the field, and have further stimulated interest in CE$\nu$NS as a probe for beyond the Standard-Model (BSM) physics~\cite{Abdullah:2022zue}. The field is thriving with several planned or ongoing experiments aiming to achieve larger exposures while reducing further the detection thresholds and the corresponding backgrounds. Such examples include the proposed \cevns detectors at the European Spallation Source (ESS) and the  Coherent Captain Mills experiments at Los Alamos~\cite{CCM:2021leg}, both  planning to exploit stopped-pion neutrinos. Further experiments which are currently under development, including  $\nu$Gen~\cite{nGeN:2022uje}, NUCLEUS~\cite{NUCLEUS:2019igx}, RICOCHET~\cite{Billard:2016giu}, MINER~\cite{MINER:2016igy}, $\nu$IOLETA~\cite{Fernandez-Moroni:2020yyl}, TEXONO~\cite{TEXONO:2024vfk}, NUXE~\cite{Ni:2021mwa}, CHILLAX~\cite{Bernard:2022zyf}, RED-100~\cite{Akimov:2024lnl}, Scintillating Bubble Chamber~\cite{SBC:2021yal} etc. are expected to bring new light in low-threshold reactor antineutrino-induced neutrino-nucleus scattering processes.

In view of the large statistics of neutrino-nucleus events anticipated at the next generation experiments, we are motivated to explore subdominant effects such as those coming from inelastic neutral-current (NC) neutrino-nucleus scattering. To achieve a realistic and accurate description of the nuclear spectra required for such high energy scattering, we utilize a hybrid model combining the nuclear shell model (NSM) and the microscopic quasiparticle-phonon model (MQPM) for nuclear structure calculations. The main advantage of this method is that it enables using the best features of both models by adopting them to model different parts of the nuclear spectra. Similar hybrid approaches in scattering calculations have been previously explored~\cite{Kolbe:1999vc, Juodagalvis:2004pj, Dzhioev:2016hvy}, where the NSM was used for allowed transitions and the random-phase approximation (RPA) for forbidden ones. Here we present a novel quasiparticle-based hybrid approach, which aims to serve as a powerful tool in inelastic neutrino-nucleus scattering calculations.

Building upon our previous work~\cite{Papoulias:2024vdc}, where we introduced a novel formalism for computing the scattering cross section in terms of the nuclear recoil energy, here we extend the nuclear calculations to higher incoming neutrino energies as well as antineutrinos. As emphasized in our previous work, from the experimental point of view the nuclear recoil energy is the main observable in neutrino scattering experiments and cannot be neglected if realistic detector simulations are to be performed. Here we report cross sections in terms of the incoming lepton energy for up to 100 MeV, folded cross sections for stopped-pion neutrinos, and present results for the event rates as functions of the nuclear recoil energy. The nuclear targets considered are the stable thallium isotopes $^{203/205}$Tl, which are present as dopants in a number of past and currently operating neutrino and dark matter detectors \cite{COSINE-100:2019lgn, DAMA:2008jlt, PICO-LON:2015rtu, Amare:2019jul, SABRE:2018lfp}, and have potential for solar neutrino scattering experiments in their own right~\cite{Kostensalo:2019cua}. Beyond the energy range of the solar neutrino spectrum, dark matter direct detection experiments are also sensitive to another source of astrophysical neutrinos, i.e., those coming from  diffuse supernova background (DSNB) neutrinos~\cite{Beacom:2010kk}.  Similarly to solar neutrinos, DSNB also forms an irreducible source of background in dark matter direct detection experiments~\cite{AristizabalSierra:2021kht, OHare:2021utq}. For the latter, we show that the adopted hybrid nuclear  model is  optimized for inelastic neutrino-nucleus scattering calculations. We find that DSNB neutrinos are of particular interest since the inelastic and \cevns rate can be of comparable size.

\section{Theory}
The NSM has the advantage of typically reproducing the low-energy spectrum of a nucleus with excellent agreement when compared to experimental data. This is especially important in low-energy neutrino scattering, such as with solar neutrinos, where contributions are dominated by a small number of low-lying states. Accurate modeling of these states is also important in higher energy scattering.  While their overall significance decreases as the neutrino energy increases, they still represent a notable fraction of the total scattering cross section. This is especially true when the lowest excited states can be reached from the ground state via an allowed transition.

On the other hand, the NSM has the disadvantage of being computationally demanding or even inadequate when modeling high-lying states that are important in higher-energy neutrino scattering, such as e.g. supernova or stopped-pion neutrinos. The computations required for obtaining a realistic excitation spectrum for the description of such scattering can be timewise prohibitively expensive, especially for open-shell nuclei with large numbers of valence nucleons. The NSM is also wholly incapable of modeling certain high-lying giant resonances which contribute substantially to scattering at these energies.

The MQPM is a quasiparticle-based nuclear model for the description of odd-$A$ nuclei. Contrary to the NSM, it tends to excel at modeling the nuclear spectrum at higher energies relevant for supernova and stopped-pion neutrinos, but leaves room for improvement at the low-energy end of the spectrum. It can describe states with energies of 25 MeV and above in reasonable computational times, and it performs well in modelling the overall features of the spectrum. This is achieved by virtue of considerably larger valence spaces compared to typical NSM calculations, often encompassing several major shells. We provide a brief review of the idea and mathematical formalism of the MQPM below.

MQPM states are built on top of an even-even reference nucleus state by using the excitation creation operator \cite{Toivanen1995, Toivanen:1998zz}
\begin{equation}
    \Gamma^{\dagger}(JM)_i = \sum_{n}C_{n}^{i}a_{nJM}^{\dagger} + \sum_{b\omega}D_{b\omega}^{i}\left[a_{b}^{\dagger}Q_{\omega}^{\dagger}\right]_{JM},
\end{equation}
where $a_{\alpha}^{\dagger}$ is the BCS quasiparticle creation operator and $Q_{\omega}^{\dagger}$ the quasiparticle random-phase approximation (QRPA) phonon creation operator \cite{Suh2007}
\begin{equation}
    Q_{\omega}^{\dagger} \equiv Q_{J_\omega\pi_\omega k_\omega}^{\dagger} = \sum_{a\leq b}\left[X^{\omega}_{ab}A_{ab}^{\dagger}(JM)-Y^{\omega}_{ab}\tilde{A}_{ab}(JM)\right].
\end{equation}
The quantum numbers $\alpha$ of a quasiparticle are given in Baranger notation \cite{Baranger:1960qge}, the quantum numbers $J_\omega(M)$, $\pi_\omega$ and $k_\omega$ of a phonon are the angular momentum (projection), parity and enumeration quantum numbers respectively. The quasiparticle operators read
\begin{equation}
\begin{aligned}
        A_{ab}^{\dagger}(JM) \equiv & \frac{\sqrt{1+\delta_{ab}(-1)^{J}}}{1+\delta_{ab}}\left[a_{a}^{\dagger}a_{b}^{\dagger}\right]_{JM} \, , \\
    \tilde{A}_{ab}(JM) =& (-1)^{J+M}A_{ab}(J-M) \, .
    \end{aligned}
\end{equation}
The MQPM states, therefore, contain both quasiparticle and quasiparticle-phonon contributions, which represent one- and three-quasiparticle configurations respectively.  In this regard the MQPM is more limited than the NSM, as although the NSM valence space can often be quite small, all the configurations representing Slater determinants built from the single-particle orbitals within  said valence space are usually included in the calculations. In contrast, higher order configurations (e.g. quasiparticle-phonon-phonon representing five-quasiparticle configurations) are not included in our MQPM calculations. One- and three-quasiparticle configurations represent the most prominent contributions to nuclear wave-functions, and thus the agreement with experiment for the lowest states is usually decent, but rather poor compared to that of the NSM. The spin-parities of the MQPM ground state and the lowest excited states are typically correct, but their energies and order with respect to each other may be off.

The procedure of applying the MQPM in practice first requires the construction of a single-particle basis $\{c_{\alpha}^{\dagger}\}$ from which the BCS quasiparticles are obtained according to the Bogolyubov-Valatin transform
\begin{equation}
    a_{\alpha}^{\dagger} = u_ac_{\alpha}^{\dagger} + v_a\tilde{c}_{\alpha}, \quad \tilde{a}_{\alpha} = u_a\tilde{c}_{\alpha} - v_ac_{\alpha}^{\dagger}.
\end{equation}
The occupation and vacancy amplitudes $v_a$ and $u_a$ are obtained by solving the BCS equations discussed in detail in e.g. \cite{Suh2007}. After this, the amplitudes $\mathsf{X}^{\omega}$ and $\mathsf{Y}^{\omega}$, and the amplitudes $C^{i}$ and $D^{i}$ can be obtained by solving the QRPA and MQPM equations respectively. We used a Coulomb-corrected Woods-Saxon single-particle basis with Bohr-Mottelson parametrization \cite{Bohr:1969}, and for the BCS and QRPA calculations we chose the even-even reference nuclei $^{204/206}$Pb for $^{203/205}$Tl respectively. The BCS calculations were performed by fitting the energy of the lowest neutron quasiparticle to the experimental pairing gap $\Delta_{\textrm{n}}$ for neutrons. The reference nuclei are proton-magic, so we performed the BCS calculations for $^{206/208}$Po fitting the lowest proton quasiparticle energies to the pairing gaps $\Delta_{\textrm{p}}$, and then used the same fitting parameters for protons in the $^{204/206}$Pb BCS calculations respectively.

In the QRPA stage we fitted the phenomenological particle-particle and particle-hole strength parameters $\mathsf{g}_{\mathsf{pp}}$ and $\mathsf{g}_{\mathsf{ph}}$ so that the lowest states of natural parity matched their experimental counterparts in energy. The MQPM calculations were then performed using the same values for the parameters as in the QRPA calculations. The NSM calculations were identical to our previous work \cite{Papoulias:2024vdc} where we utilized the computer code NuShellX@MSU \cite{Brown:2014bhl} using the model space jj56pn with the interaction $khhe$. We combined the models by using NSM states up to 3 MeV and MQPM states above this threshold for computing the NC one-body transition densities between the excited states and the ground state, which are needed for computing the nuclear matrix elements of the scattering operators.

\begin{figure}
\centering
\resizebox{\columnwidth}{!}{
\includegraphics{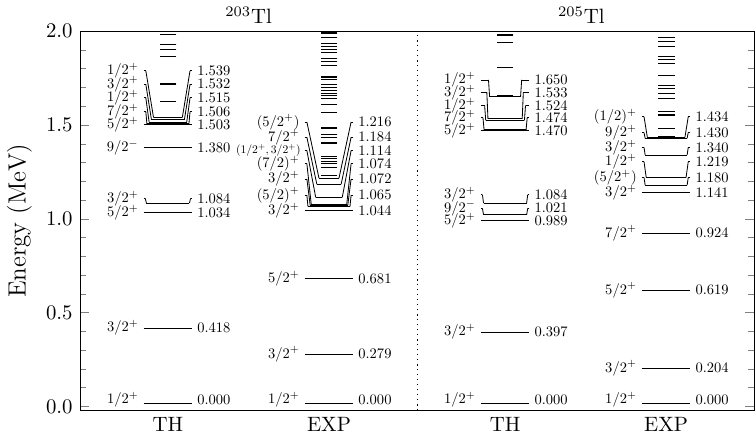}
}
\caption{Lowest states in the nuclear spectra obtained by using MQPM compared to experimental data \cite{Kondev:2005bkq, Kondev:2020yee}. The agreement with experiment is fair, but inferior to pure NSM spectra that we utilized in an earlier work~\cite{Papoulias:2024vdc}.}
\label{fig:spectra}
\end{figure}

The nuclear spectra obtained with the MQPM are illustrated in Fig.~\ref{fig:spectra}. The angular momentum and parity of the ground states and the lowest excited states are reproduced correctly for both nuclei, but the energies of the excitations are slightly off, as is often the case with MQPM \cite{Hellgren:2022yvo, Hellgren:2024gij}. While the agreement with experiment is fair for both nuclei, the NSM performs noticeably better in the low-energy end of the spectra.

Semileptonic nuclear processes can be described using the Donnelly-Walecka formalism, in which inelastic NC neutrino-nucleus scattering to the lowest order can be modeled by a current-current  effective Hamiltonian between the lepton ($j_{\mu}(\textbf{x})$) and the hadron ($\mathcal{J}_{\mu}(\textbf{x})$) currents~\cite{Donnelly:1978tz, Walecka:2004}
\begin{equation}
    \hat{H}_{\textrm{eff}} = \frac{G}{\sqrt{2}}\int\textrm{d}^3\textbf{x}j_{\mu}(\textbf{x})\mathcal{J}^{\mu}(\textbf{x}),
\end{equation}
where $G\equiv G_F = 1.1664\times10^{-5}$ GeV$^{-2}$ is the Fermi constant. The double-differential scattering cross section from an initial state $\ket{i}$ to a final state $\ket{f}$ is then obtained by decomposing the matrix elements $|\bra{f}\hat{H}_{\textrm{eff}}\ket{i}|^2$ in terms of spherical tensor operators, and can be shown to be~\cite{Walecka:2004,Ydrefors2011a}
\begin{equation}
    \frac{\textrm{d}\sigma_{i\rightarrow f}}{\textrm{d}\Omega\textrm{d}\omega} = \frac{G^2|\textbf{k}'|E_{\nu'}}{\pi(2J_i+1)}\left(\sum_{J\geq 0}\sigma^{J}_{\rm CL} + \sum_{J\geq 1}\sigma^{J}_{\rm T}\right),
\end{equation}
where $\Omega$, $E_{\nu'}$ and $\textbf{k}'$ are the angular coordinates, energy and 3-momentum of the scattered neutrino, $\omega$ the energy of the final nuclear state, $J_i$ the angular momentum of the initial nuclear state and $\sigma^{J}_{\rm CL}$ and $\sigma^{J}_{\rm T}$ are the Coulomb-longitudinal and transverse contributions to the cross section which are discussed in detail in \cite{Donnelly:1978tz, Walecka:2004, Ydrefors2011a}.
The latter are expressed in terms of the lepton traces $\sum_\text{spins}l_\mu l_\nu^*$ and the nuclear matrix elements of the Coulomb, Longitudinal, Transverse electric and Transverse magnetic operators $\mathcal{M, L, T^\text{el}, T^\textrm{mag}}$, respectively 
\begin{equation}
	\begin{split}
	&\sigma^J_{\textrm{CL}} = l_0l_0^* |(J_f||\mathcal{M}_J(q)||J_i)|^2 + l_3l_3^*|(J_f||\mathcal{L}_J(q)||J_i)|^2 \\& -2\textrm{Re}\,  l_3l_0^*\left[(J_f||\mathcal{L}_J(q)||J_i)(J_f||\mathcal{M}_J(q)||J_i)^*\right] \, ,
	\end{split}
\end{equation}
\begin{equation}
	\begin{split}
	&\sigma^J_{\textrm{T}} = \frac{\mathbf{l}\cdot \mathbf{l}^* - l_3l_3^*}{2} \left[|(J_f||\mathcal{T}^{\textrm{el}}_J(q)||J_i)|^2 + |(J_f||\mathcal{T}^{\textrm{mag}}_J(q)||J_i)|^2\right] \\& - 2 i\textrm{Re}\,  \frac{\mathbf{l}\times 
    \mathbf{l}^*}{2} \left[(J_f||\mathcal{T}^{\textrm{mag}}_J(q)||J_i)(J_f||\mathcal{T}^{\textrm{el}}_J(q)||J_i)^*\right] \, .
	\end{split}
\end{equation}
 The lepton traces given in terms of the incident neutrino energy, nuclear excitation energy and recoil energy ($T$) are~\cite{Papoulias:2024vdc}
\begin{equation}
  \sum_\mathrm{spins} l_0 l_0^* =  \frac{4 E_\nu^2-4 E_\nu (T+\omega )-2 m_\mathcal{N} T+\omega  (2 T+\omega )}{2
   E_\nu (E_\nu-\omega )} \, ,
\end{equation}
\begin{equation}
  \sum_\mathrm{spins}  l_3 l_0^* = \frac{T+\omega}{\sqrt{2 m_\mathcal{N} T}} \,  l_0l_0^* \, ,
\end{equation}
\begin{equation}
  \sum_\mathrm{spins} l_3l_3^* =  \frac{(T+\omega)^2}{2 m_\mathcal{N} T} \,  l_0l_0^* \, ,
\end{equation}
\begin{equation}
  \sum_\mathrm{spins} (\mathbf{l} \cdot \mathbf{l^*} - l_3 l_3^*) =  \left(1-\frac{\omega  (2 T+\omega )}{2 m_\mathcal{N} T}\right) \left(l_0l_0^* + \frac{2 m_\mathcal{N}
   T}{E_{\nu } \left(E_{\nu }-\omega \right)}\right) \, ,
\end{equation}
\begin{equation}
\sum_\mathrm{spins}   \frac{-i}{2}(\mathbf{l} \times \mathbf{l^*})_3 =  \frac{\left(2 E_{\nu } - \omega \right) (2 m_\mathcal{N} T-\omega  (2 T+\omega ))}{2
    E_{\nu } \left(E_{\nu } - \omega\right) \sqrt{2 m_\mathcal{N} T}}  \, ,
\end{equation}
assuming $T\ll E_\nu $ and $T\ll m_\mathcal{N}$, with $m_\mathcal{N}$ being the nuclear mass. The differential cross section in terms of the nuclear recoil energy is then given by~\cite{Papoulias:2024vdc}
\begin{equation}
   \frac{\textrm{d} \sigma}{\textrm{d} T} =   \frac{\textrm{d} \sigma}{\textrm{d} \cos \theta}    \frac{m_\mathcal{N}}{E_\nu (E_\nu - \omega)} \, ,
\end{equation}
with $T \approx \left[E_\nu (E_\nu - \omega) (1-\cos \theta) + \omega^2/2\right]/m_\mathcal{N}$.

\section{Results}

\begin{table}[h]
\centering
\small
\begin{tabular}{l c c c c} 
 \hline
 &\multicolumn{2}{c}{$^{203}\textrm{Tl}$}&\multicolumn{2}{c}{$^{205}\textrm{Tl}$}\\
 $E_{\nu}$ [MeV]&$\sigma_{\nu}$&$\sigma_{\overline{\nu}}$&$\sigma_{\nu}$&$\sigma_{\overline{\nu}}$\\
 \hline
 5.0&5.099(0)&4.992(0)&5.599(0)&5.471(0)\\
 10.0&1.315(2)&1.261(2)&1.255(2)&1.209(2)\\
 15.0&8.569(2)&8.073(2)&8.351(2)&7.866(2)\\
 20.0&2.393(3)&2.196(3)&2.365(3)&2.160(3)\\
 25.0&4.952(3)&4.404(3)&4.956(3)&4.372(3)\\
 30.0&8.724(3)&7.495(3)&8.814(3)&7.497(3)\\
 35.0&1.380(4)&1.145(4)&1.403(4)&1.151(4)\\
 40.0&2.017(4)&1.617(4)&2.056(4)&1.630(4)\\
 45.0&2.772(4)&2.154(4)&2.826(4)&2.173(4)\\
 50.0&3.631(4)&2.743(4)&3.694(4)&2.765(4)\\
 55.0&4.581(4)&3.375(4)&4.645(4)&3.396(4)\\
 60.0&5.613(4)&4.043(4)&5.670(4)&4.062(4)\\
 65.0&6.718(4)&4.739(4)&6.761(4)&4.755(4)\\
 70.0&7.884(4)&5.454(4)&7.908(4)&5.467(4)\\
 75.0&9.095(4)&6.175(4)&9.097(4)&6.185(4)\\
 80.0&1.033(5)&6.886(4)&1.031(5)&6.895(4)\\
 85.0&1.156(5)&7.573(4)&1.152(5)&7.581(4)\\
 90.0&1.277(5)&8.224(4)&1.271(5)&8.231(4)\\
 95.0&1.394(5)&8.832(4)&1.384(5)&8.837(4)\\
 100.0&1.503(5)&9.392(4)&1.492(5)&9.394(4)\\
 \hline
\end{tabular}
\caption{Inelastic scattering cross section off $^{203/205}$Tl as a function of the incoming neutrino energy $E_{\nu}$. 
The data is given in the format $B(p)$, and cross sections are obtained by $\sigma = B\times10^p\times D$, where the units are $D = 10^{-44}$ cm$^2$.
}
\label{tab:xsect-range-high}
\end{table}
\begin{table}[h]
\centering
\small
\begin{tabular}{l c c c c} 
 \hline
 &\multicolumn{2}{c}{$^{203}\textrm{Tl}$}&\multicolumn{2}{c}{$^{205}\textrm{Tl}$}\\
 $E_{\nu}$ [MeV]&$\sigma_{\nu}$&$\sigma_{\overline{\nu}}$&$\sigma_{\nu}$&$\sigma_{\overline{\nu}}$\\
 \hline
 1.0&3.109(1)&3.023(1)&6.942(0)&7.452(0)\\
 2.0&2.180(4)&2.162(4)&6.994(4)&6.933(4)\\
 3.0&3.713(5)&3.669(5)&7.225(5)&7.133(5)\\
 4.0&1.945(6)&1.913(6)&2.514(6)&2.470(6)\\
 5.0&5.099(6)&4.992(6)&5.599(6)&5.471(6)\\
 6.0&1.011(7)&9.836(6)&1.013(7)&9.856(6)\\
 7.0&1.743(7)&1.681(7)&1.647(7)&1.596(7)\\
 8.0&3.159(7)&3.030(7)&2.941(7)&2.840(7)\\
 9.0&6.888(7)&6.609(7)&6.503(7)&6.275(7)\\
 10.0&1.315(8)&1.261(8)&1.255(8)&1.209(8)\\
 11.0&2.204(8)&2.109(8)&2.116(8)&2.032(8)\\
 12.0&3.362(8)&3.207(8)&3.244(8)&3.102(8)\\
 13.0&4.801(8)&4.562(8)&4.649(8)&4.425(8)\\
 14.0&6.531(8)&6.181(8)&6.346(8)&6.010(8)\\
 15.0&8.569(8)&8.073(8)&8.351(8)&7.866(8)\\
 16.0&1.093(9)&1.025(9)&1.068(9)&1.000(9)\\
 17.0&1.363(9)&1.271(9)&1.336(9)&1.244(9)\\
 18.0&1.668(9)&1.548(9)&1.640(9)&1.517(9)\\
 19.0&2.011(9)&1.856(9)&1.982(9)&1.822(9)\\
 20.0&2.393(9)&2.196(9)&2.365(9)&2.160(9)\\
 \hline
\end{tabular}
\caption{Same as Table~\ref{tab:xsect-range-high}, but for a lower incoming neutrino energy range and with $D = 10^{-50}$ cm$^2$.}
\label{tab:xsect-range-low}
\end{table}

\begin{figure*}[h]
\centering
\includegraphics[width=0.4\textwidth]{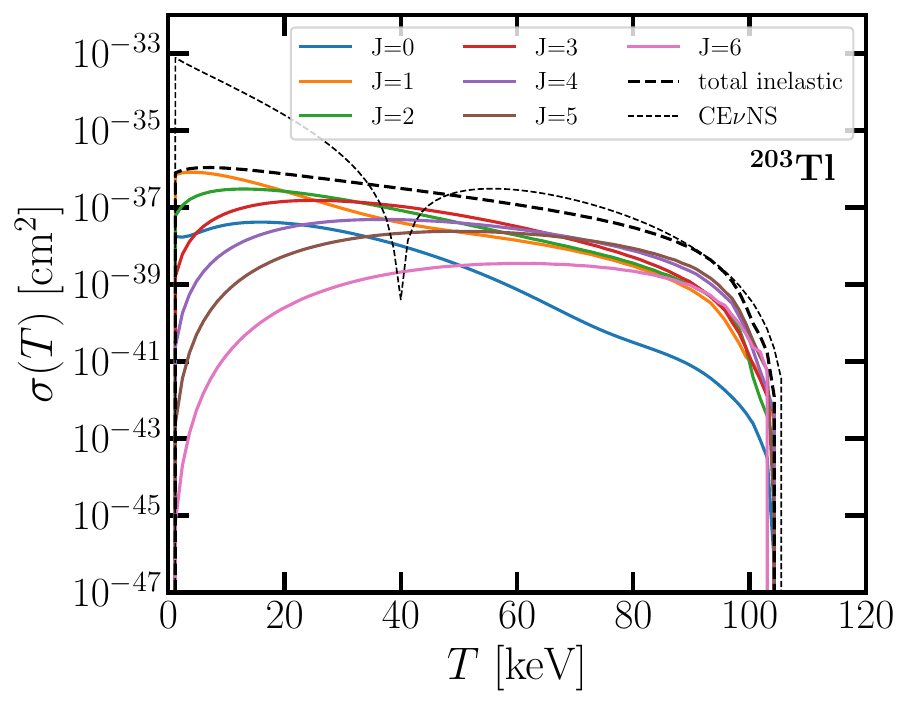}
\hspace{1cm}
\includegraphics[width=0.4\textwidth]{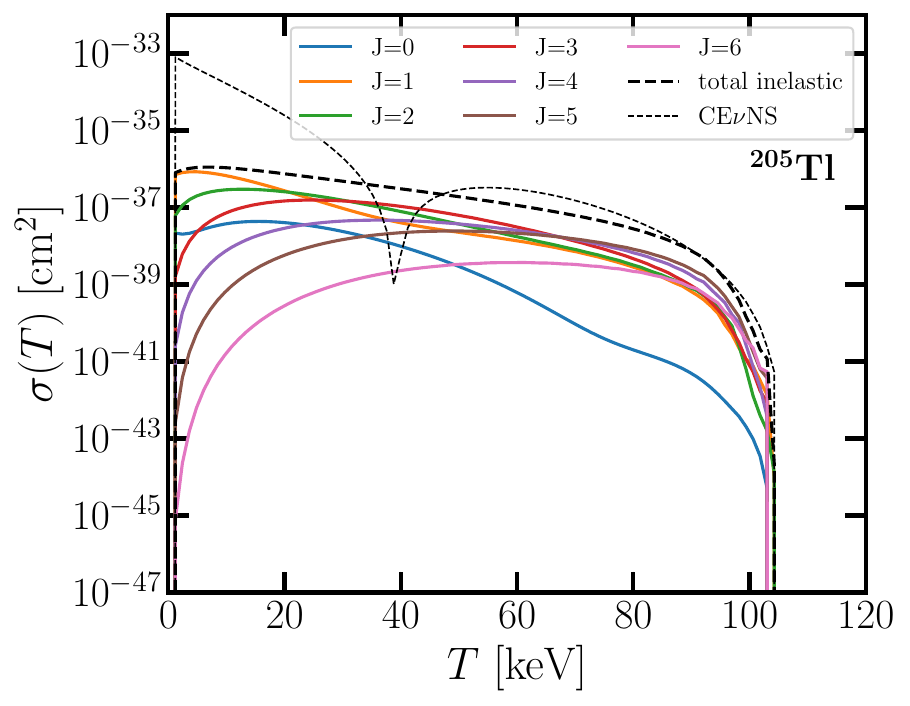}
\caption{Integrated inelastic neutrino-nucleus scattering cross sections as a function of the nuclear recoil energy, for $^{203}$Tl (left) and $^{205}$Tl (right). The results are given for the individual $J$-transitions and their sum. The corresponding \cevns cross sections are also given for comparison.}
\label{fig:dsdT}
\end{figure*}

In Fig.~\ref{fig:dsdT} we illustrate the scattering cross sections as a function of the nuclear recoil energy, for $^{203}$Tl (left panel) and $^{205}$Tl (right panel). The results are given for the individual  $J$-transitions as well as their sum. A comparison with the corresponding 
\cevns cross section is also given for each isotope, where one can see that at around $T\simeq 40$~keV, the inelastic cross section dominates. For both nuclei  the $J = 1 $ transition
is the most relevant up to about  $T \sim  10$~keV, while in the range $20\lesssim T \lesssim 70$~keV   $J = 2$ and $J=3$ dominate, and finally, for higher recoil energies, $J=4$ and $J=5$ become the dominant ones. While this is the same general behavior observed in Ref.~\cite{Papoulias:2024vdc} based solely on NSM calculations, the  aforementioned recoil ranges were slightly different.  

The scattering cross sections as functions of the incoming neutrino energy for both neutrinos and antineutrinos are presented in Tables \ref{tab:xsect-range-high} and \ref{tab:xsect-range-low}.
Comparing the results in the latter table for neutrinos to our previous work \cite{Papoulias:2024vdc} where we used solely the NSM, we clearly see that after the 3~MeV hybrid threshold the cross sections deviate slightly from one another, until at 9~MeV the new hybrid model results are already over 1.5/nearly 2 times larger than the pure NSM results for $^{203/205}$Tl. At 20 MeV the results differ by an order of magnitude. This has no effect on the cross sections of the lower energy solar neutrinos that have $Q$-values below the hybrid threshold, but $hep$, and $^{8}$B neutrino cross sections are affected.
\begin{figure*}
\centering
\includegraphics[width=0.325\textwidth]{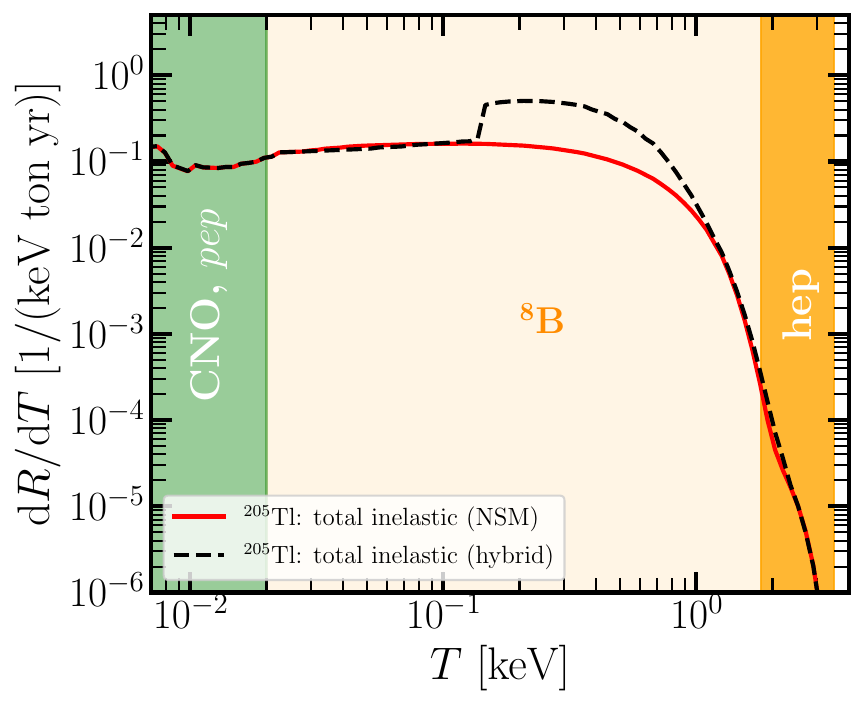}
\includegraphics[width=0.34\textwidth]{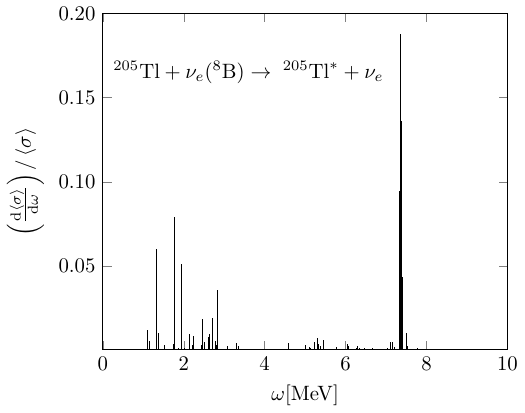}
\includegraphics[width=0.325\textwidth]{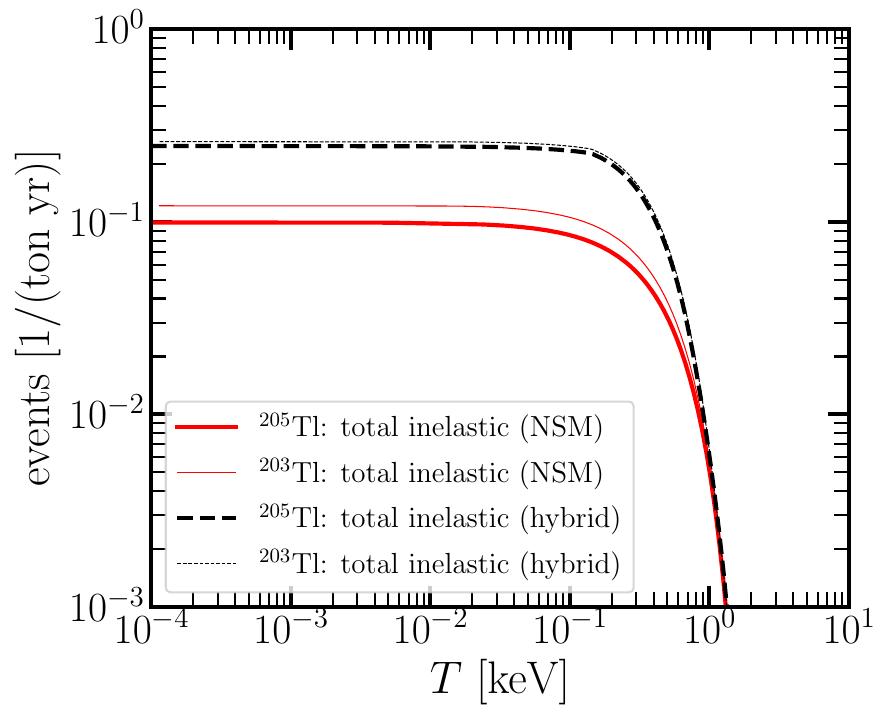}
\caption{Differential (left)  and integrated (right) number of events above threshold for solar neutrino scattering off $^{203/205}$Tl using the hybrid model compared to our previously obtained pure NSM result. The center panel shows the contributions to the total folded $^{8}$B solar-neutrino scattering cross section off $^{205}$Tl from individual final nuclear states with excitation energy $\omega$, normalized to the total folded cross section.}
\label{fig:comparison_solar}
\end{figure*}

In the left panel of Fig.~\ref{fig:comparison_solar} we present the differential event rates for solar neutrinos, comparing the new results of the hybrid model with those based on pure NSM calculations~\cite{Papoulias:2024vdc}. The flux normalizations of the solar neutrino spectra are taken from~\cite{Baxter:2021pqo}. The rates of $^{203}$Tl and $^{205}$Tl are rather similar, and thus, only the latter is presented to avoid overcrowding the plot. As expected, the results corresponding to the low-energy sources such as CNO and $pep$ neutrinos are identical for both models. On the other hand, in the energy region where $^8$B and $hep$ neutrinos dominate, a notable difference between the NSM and hybrid model  predictions is found. This is particularly pronounced  for nuclear recoil energies in the range $0.1 \lesssim T \lesssim 1$~keV. On the other hand, this discrepancy is not present for the high-energy tail of the $hep$ neutrino spectrum that lies entirely above the $^{8}$B spectrum.

The discrepancy can be explained by looking at contributions to the total folded cross section from individual final nuclear states. These contributions are illustrated for the scattering of $^{8}$B neutrinos off $^{205}$Tl in the center panel of Fig.~\ref{fig:comparison_solar}. While it can be clearly seen that there are a number of highly-contributing sub-hybrid threshold states further highlighting the importance of accurate modelling of these states using the NSM, over 50\% of the contributions come from a small number of states centered narrowly at around 7.4~MeV and having spin-parities of $1/2^+$ or $3/2^+$. This is likely a spin-flip $M$1 giant resonance which are known to contribute substantially in NC neutrino scattering~\cite{Hellgren:2022yvo}. The latter was entirely absent in our previous calculations due to the limitations of the NSM.

In~\cite{Papoulias:2024vdc} we computed the limits for the recoil energy as functions of the nuclear excitation energy and incoming neutrino energy. The lower limit of $T$ for the excitation energy of the resonance is about $0.14$ keV, which coincides with the large jump in the differential event rate of the hybrid model. Similarly, higher nuclear excitation energies have higher upper limits for $T$, and we see the effect of the resonance diminishes as $T$ grows. For a neutrino energy corresponding to the $Q$-value of the $hep$ neutrino producing reaction the upper limit for the resonance excitation energy is about $2.4$ keV\footnote{Coincidentally, this is also the absolute upper limit of $T$ for $^{8}$B neutrinos when the nuclear excitation energy is set to zero.}, by which point only the states below this energy contribute, and the contributions come almost entirely from low-energy NSM states, and the differential number of events for both models coincides.

It can also be clearly seen that the inclusion of the higher energy states greatly affects the event rate. Thus, even at such low incoming neutrino energies the hybrid model is needed for a realistic description of the scattering. The corresponding  integrated number of events  is illustrated in the right panel of Fig.~\ref{fig:comparison_solar} along with the previously reported result for which only the NSM was used. In this case the integrated rates are plotted for both $^{203/205}$Tl isotopes. It is noteworthy that, between the new hybrid model results and the previous NSM-based ones,  the former are found to be enhanced by factor 2 or more. Let us stress, however,  that when comparing the inelastic rates presented here with the corresponding \cevns ones, the latter are larger by several orders of magnitude and hence they are not shown (for such a comparison, see Ref.~\cite{Papoulias:2024vdc}).

Next, the new hybrid nuclear model is  applied for computing inelastic neutrino-nucleus scattering cross sections involving stopped-pion neutrinos~\cite{Louis:2009zza}. Stopped-pion neutrinos are considerably more energetic than solar neutrinos extending up to $\sim53$~MeV, thus requiring an accurate description of the nucleus up to considerably higher excitation energies, which the hybrid model is well-suited for. To get an idea of the order of magnitude of these cross sections, we first present the folded cross sections $\overline{\sigma}$ for stopped-pion neutrinos in Table~\ref{tab:folded-xsects}. The folded cross sections are almost identical for both nuclei which is expected as their spectra are quite similar. It is only at very low neutrino energies when the differences in the energies and wave-functions of the low-energy states of the two nuclei are so noticeable, that the cross section differs by an order of magnitude. This can be also inferred from Table~\ref{tab:xsect-range-low}.
\begin{figure*}[t]
\centering
\includegraphics[width=0.4\textwidth]{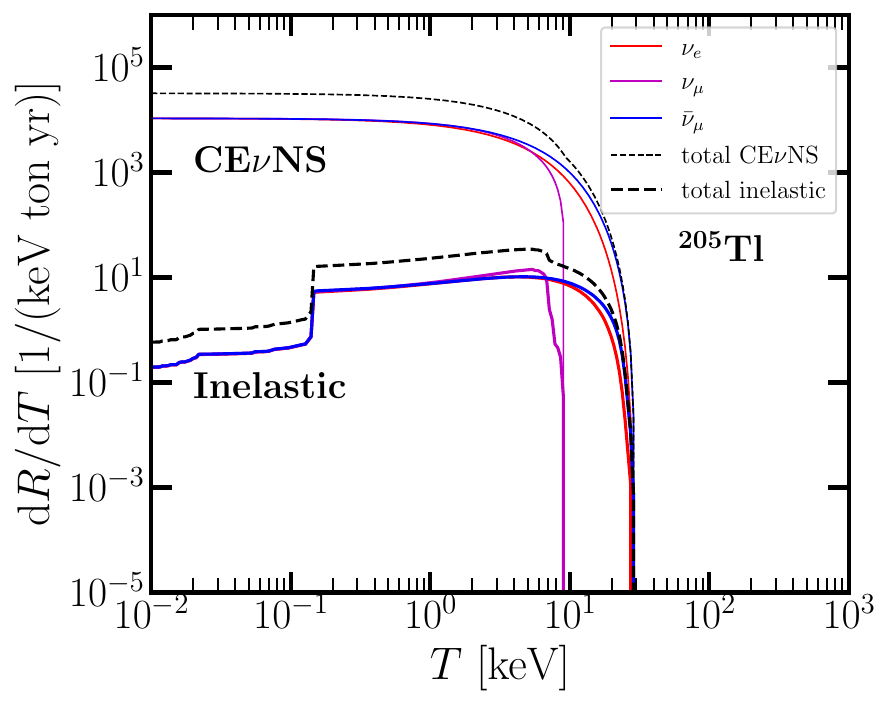}
\hspace{1cm}
\includegraphics[width=0.4\textwidth]{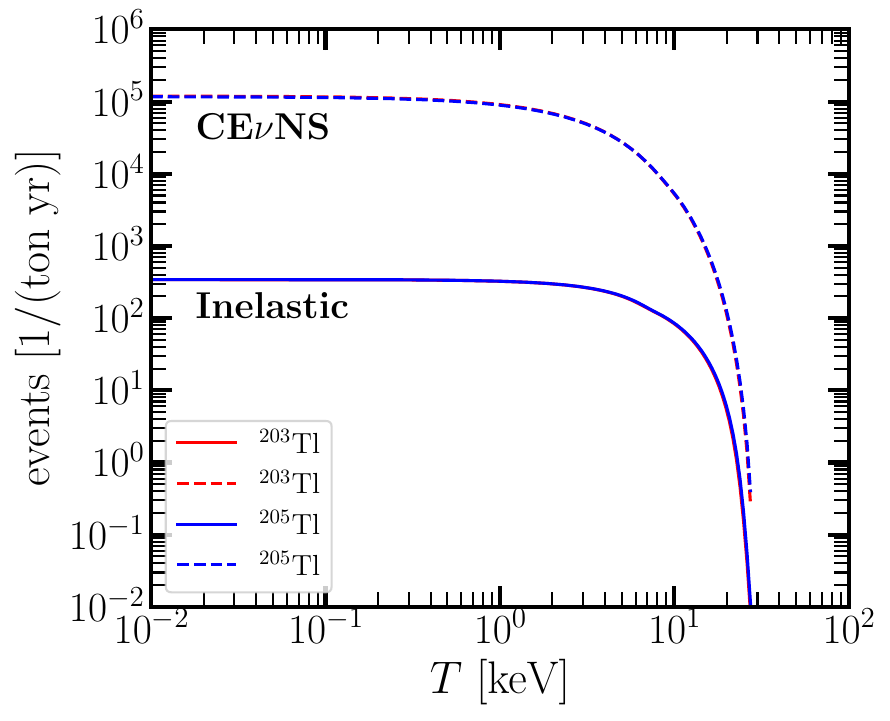}
\caption{Differential (left) and integrated (right) number of events above threshold for inelastic stopped-pion neutrino scattering off $^{203/205}$Tl nuclei using the hybrid model. The results are given for the individual neutrino flavors and their sum. The corresponding \cevns spectra are also shown for comparison.}
\label{fig:stopped-pion}
\end{figure*}
\begin{table}
\centering
\small
\begin{tabular}{l c c} 
 \hline
 Neutrino&$\overline{\sigma}_{^{203}\rm Tl}$&$\overline{\sigma}_{^{205}\rm Tl}$\\
  \hline
 $\nu_{e}$&1.306&1.326\\
 $\nu_{\mu}$&0.855&0.863\\
 $\overline{\nu}_{\mu}$&1.502&1.512\\
 \hline
\end{tabular}
\caption{Total folded inelastic neutrino-nucleus scattering cross sections for stopped-pion neutrinos. The results are given in units of $10^{-40}~\mathrm{cm^2}$.}
\label{tab:folded-xsects}
\end{table}
\begin{figure*}[h]
\centering
\includegraphics[width=0.4\textwidth]{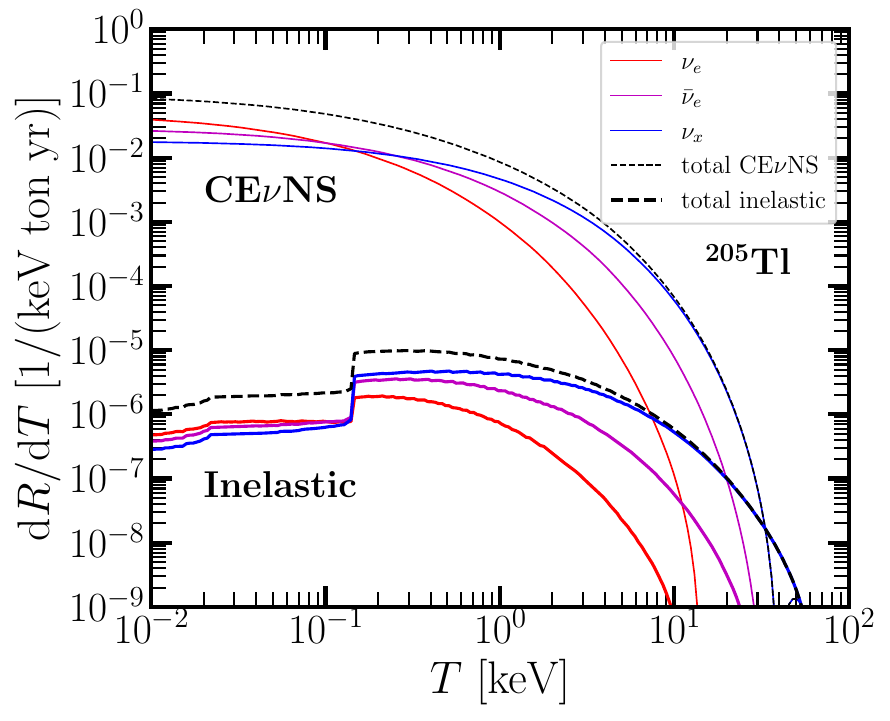}
\hspace{1cm}
\includegraphics[width=0.4\textwidth]{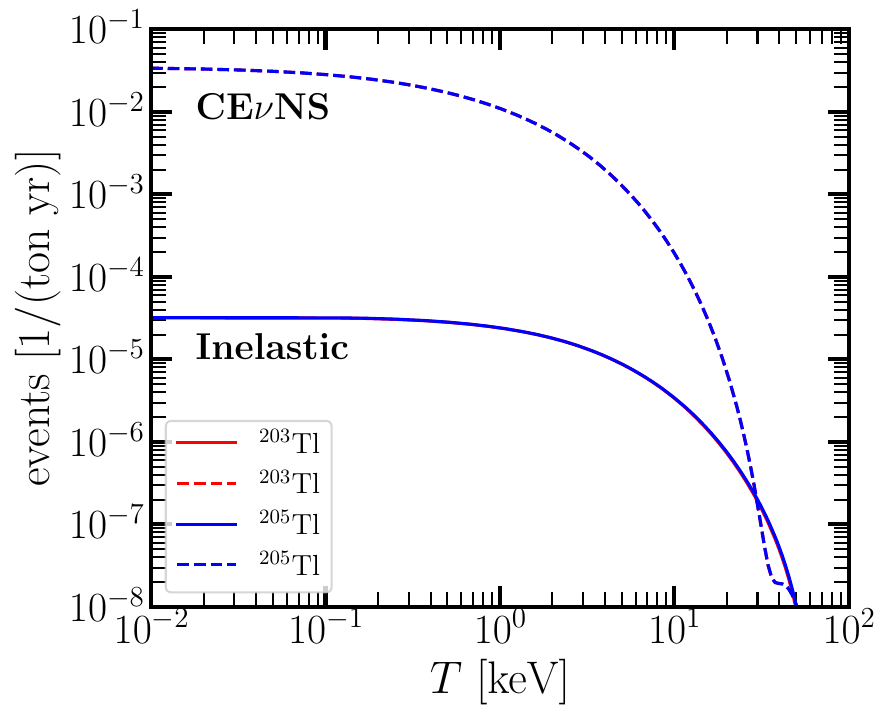}
\caption{Same as Fig.~\ref{fig:stopped-pion}, but for the case of DSNB neutrinos.}
\label{fig:DSNB}
\end{figure*}

In Fig.~\ref{fig:stopped-pion} we present the differential (left panel) and integrated (right panel) event rates assuming stopped-pion neutrinos, for both thallium isotopes. The calculation assumes a typical flux of $10^7 \nu/(\mathrm{cm^2 \, s})$, which is expected at spallation sources. The differential event rates in the left panel are given for each flavor separately as well as for their sum. The integrated events shown in the right panel are summed over neutrino flavors.  Similarly to the case of solar neutrinos, the two isotopes yield very similar rates. We can also compare the total number of events in inelastic scattering to CE$\nu$NS. It becomes evident that, while the inelastic channel is still clearly subdominant as was the case for solar neutrinos, the difference between them is orders of magnitude smaller. This is partly due to the more accurate hybrid-model treatment of the nuclear structure, but also due to the higher energy of stopped-pion neutrinos. It is also  interesting to notice that in the high energy tail of the event spectrum the two channels are becoming competitive, which implies that the high energy \cevns spectra must be treated with special care in phenomenological BSM analyses.

The previous discussion implies that the more energetic a neutrino source is, the more competitive the inelastic channel becomes versus the dominant \cevns one (see also Fig.~\ref{fig:dsdT}). With this in mind, we proceed by computing the expected event rates induced by DSNB neutrinos. DSNB consists of all neutrino flavors, with the three different components accounting for $\nu_e$, $\bar{\nu}_e$ and $\nu_x$ ($x$ denotes the remaining neutrino and antineutrino flavors). The distributions are commonly parametrized using Fermi-Dirac or power-law spectra, and characterized by temperatures of 3~MeV for $\nu_e$, 5~MeV for $\bar{\nu}_e$, and 8~MeV for $\nu_x$~\cite{Beacom:2010kk}. The hybrid model considered here, is thus fully suitable for accurately describing the inelastic scattering channel. In the left and right panels of Fig.~\ref{fig:DSNB} we present the corresponding differential and integrated event rates. As previously, the differential rates are illustrated for the individual neutrino flavor contributions and their sum. To our knowledge, inelastic neutrino-nucleus event rates due to the DSNB are presented for the first time in this work.  It is worth noticing that the inelastic channel corresponding  to the $\nu_x$ component dominates over the \cevns one for $T\gtrsim 20~\mathrm{keV}
$. We thus arrive at the conclusion that if an accurate treatment of the neutrino floor is to be made, the inelastic channel must be incorporated in the calculation. Before closing this discussion we must note that since the DSNB spectra are characterized by a very low intensity flux, in order to achieve detectable rates a very large exposure is required. From the experimental point of view this is rather challenging as it seems not be possible even with the next generation dark matter direct detection experiments with multi-ton mass such as ARGO~\cite{Billard:2016giu} and DARWIN~\cite{DARWIN:2016hyl}.

\section{Conclusions}
In this paper we have studied neutral-current inelastic neutrino-nucleus scattering off the stable thallium isotopes. We have computed the total cross sections as functions of the incoming lepton energy and presented results for folded stopped-pion cross sections. For the nuclear structure calculations we have introduced a novel hybrid approach, combining the nuclear shell model and the microscopic quasiparticle-phonon model for a more realistic and accurate treatment of the nuclear spectra needed for higher-energy neutrino sources.

We have also considered the cross section as a function of the nuclear recoil energy, and computed the total number of events as function of the detector threshold for stopped-pion and solar neutrinos. The inelastic stopped-pion scattering event rates were clearly subdominant to CE$\nu$NS, but this difference between them was orders of magnitude smaller than for solar neutrinos. In connection to dark matter direct detection searches, we also compared our previously obtained shell-model based solar neutrino results to our new hybrid model results, and found that the inclusion of the high-energy MQPM states remarkably affected the number of events. Finally, we have computed the inelastic scattering rates due to DSNB neutrinos for the first time in this work. In this case we found that the inelastic rates can even surpass the \cevns ones. However, the magnitude of the expected rates would require unrealistically huge exposures.

The results indicate that the hybrid approach can be a powerful tool in future neutrino scattering calculations where good agreement with experiment is desired at the low-end of the nuclear spectrum, but where incoming lepton energies are too high for a pure NSM description of the nucleus. We have plans to extend the hybrid approach in future studies involving charged-current neutrino scattering and possibly other semi-leptonic nuclear processes, such as muon capture. Indeed, the hybrid model is also not limited to NSM and MQPM, and the possibility of utilizing it in even higher energy neutrino scattering for which the MQPM alone is insufficient is also under consideration. For instance, this approach could be used  to obtain realistic inelastic rates in view of the much more energetic Fermilab neutrino beamlines at $\nu$BDX-DRIFT experiment~\cite{AristizabalSierra:2022jgg} where MQPM states would be used up to some threshold energy and states (or a continuum of states) would be used above this threshold.

\section*{Acknowledgements}
M.H. acknowledges the financial support from the Finnish Cultural Foundation. 
D.K.P. is  supported by the Spanish grants PID2023-147306NB-I00 and CEX2023-001292-S (MCIU/AEI/10.13039/501100011033), as well as CIPROM/2021/054 (Generalitat Valenciana) and CNS2023-144124 (MCIN/AEI/10.13039/501100011033 and “Next Generation EU”/PRTR).

\bibliographystyle{elsarticle-num}
\bibliography{bibliography}

\end{document}